\documentclass{PoS}
\usepackage{amsmath}
\usepackage{xspace}
\newcommand{\pythia}{P\protect\scalebox{0.8}{YTHIA}\xspace}
\newcommand{\dipsy}{\protect\scalebox{0.8}{DIPSY}\xspace}

\newcommand{\pytppp}{P\protect\scalebox{0.8}{YTHIA}8\xspace}

\newcommand{\eg}{\emph{e.g.}\ }
\newcommand{\ie}{\emph{i.e.}\ }
\newcommand{\etal}{\emph{et al.}\ }

\newcommand{\plet}[1]{\vec{#1}}
\def\text{\mathrm}
\title{Hadronisation Models and Colour Reconnection}

\ShortTitle{Hadronisation Models and Colour Reconnection}
\author{\speaker{Christian Bierlich}%
	\thanks{In collaboration with Jesper R. Christiansen, G{\"o}sta Gustafson, Leif L{\"o}nnblad (Lund University) and Andrey Tarasov (JLab). Work supported in part by the MCnetITN FP7 Marie Curie Initial Training Network, contract PITN-GA-2012-315877. LU-TP 16-36, MCnet-16-24}
       \\ Lund University\\ E-mail: \email{christian.bierlich@thep.lu.se}}

\abstract{Enhanced production of hadrons with $s$-quark content has been observed in $pp$ collisions at LHC, and earlier in collisions of heavy nuclei. We review the string hadronisation formalism and corrections from rope hadronisation and colour reconnection, corrections that takes place in such dense environments, and are able to correctly describe data. Since such corrections are very sensitive to the modelling of transverse proton structure, we investigate two such models, and compare to final states. Finally we describe how such corrections can also give a possible explanation to collective phenomena observed in small systems.}

\FullConference{XXIV International Workshop on Deep-Inelastic Scattering and Related Subjects\\
         11-15 April, 2016\\
         DESY Hamburg, Germany}

\begin{document}
Data from $pp$ collisions at the LHC, taken with minimum bias triggers, has sparked a renewed interest in hadronisation models and models of colour reconnection. Measurements from \eg CMS \cite{Khachatryan:2011tm} showed early on, that neither ratios of hadrons with strange quark content to hadrons without strange quark content, nor ratios of baryons to mesons, were correctly predicted by Monte Carlo implementations of string hadronisation \cite{Andersson:1983jt}, \eg \pythia \cite{Sjostrand:2014zea}. We will here focus on corrections to the string hadroniztion model which introduces a dependance on the space time distribution of strings. The physical picture behind this is straightforward: If a $pp$ collision consists of multiple partonic interactions (MPIs), strings can overlap in transverse space. 
That the string hadronisation model fails to predict flavour ratios correctly in dense environments, was noted in collisions with various targets already at CERN-SPS by the NA35 \cite{Bartke:1990cn} group. Based on the idea that several strings can form a multiplet, a so called "colour rope" \cite{Biro:1984cf}, with a string tension scaling as the secondary Casimir operator of the multiplet, Sorge \etal introduced colour rope formation to RQMD \cite{Sorge:1992ej}. This allowed for an overall better description of hadronic flavour content.

The idea of rope formation and hadronisation was recently adapted to $pp$ collisions \cite{Bierlich:2014xba}. After briefly introducing the models, we will introduce a simple toy model for overlaps, before finally presenting results for the full final state.

\section{Strings and ropes}
The string model for hadronisation is a dynamical model for hadronisation, based on the picture of a linear color electric field stretched between two quarks, as they move away from each other, much like a classical spring or string. When it becomes energetically favorable for the string to break into smaller strings, a new $q\bar{q}$ pair breaks the string through tunnelling with the rate:

\begin{equation}
	\label{eq:dpdpt}
	\frac{dP}{dp_\perp} \propto \kappa \exp\left(-\frac{\pi m^2_\perp}{\kappa} \right),
\end{equation}

where $m_\perp$ is the transverse mass of the quarks and $\kappa$ is the longitudinal energy density of the string, the so-called string tension, which from both experiments and lattice calculations is estimated to be $\kappa \approx 1$ GeV/fm for a $q\bar{q}$ string in vacuum. Equation (\ref{eq:dpdpt}) is used to estimate the relative amount of string breaks where strange quarks are produced, but since it is not possible to precisely determine the ratio of quark masses theoretically, the ratio $m_s/m_u$ becomes a tunable parameter\footnote{Tunes to LEP data places $m_s/m_u \approx 0.2$, in accordance with expectation.}. The exponential surpression also implies that $c$ quarks or heavier types cannot be produced by hadronisation.
Due to the non-Abelian nature of QCD, there are several different possibilities for the strings to interact. The probablity for each type of interaction can be calculated using group properties of SU(3), as indicated in figure \ref{fig:randomwalk}. Here, we start with an ordinary triplet ($q\bar{q}$) string, and consider it's overlap with another string of the same type, going left to right. We have two distinct possibilities for the interaction. Either we end in a sextet configuration or an anti-triplet one, corresponding to the well known $\plet{3} \otimes \plet{3} = \plet{6} \oplus \bar{\plet{3}}$.

\FIGURE[t]{
\centering
\includegraphics[width=0.6\linewidth,angle=0]{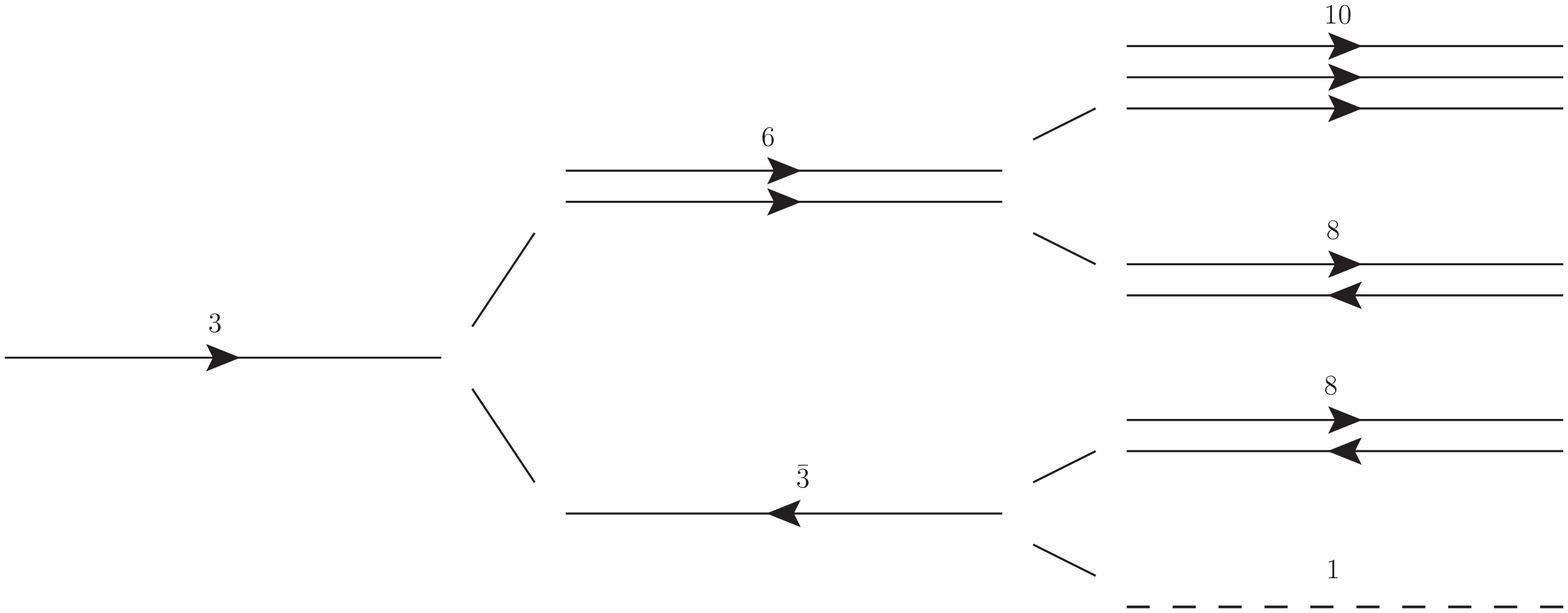}
\caption{Sketch of random walks in $q\bar{q}$ triplets, going from left to right. Furthest to the right we see the highest multiplet on the top, two junction structures, and finally a singlet.}
\label{fig:randomwalk}
}

Adding another fundamental triplet to the multiplet, takes us one more step towards the right in figure \ref{fig:randomwalk}. On the top, we have a decuplet and an octect, and on the bottom an octet and a singlet. The two octet configuration corresponds to so-called junction topologies, whereas the singlet corresponds to destructive interference between the two strings. 

\section{Consequences of higher multiplets}
There are several schemes for handling the different types of higher multiplets. The singlet configurations are often handled by minimizing some approximation of the string potential energy, the so-called $\lambda$-measure. Such a minimization is a correction to the non-perturbative hadronisation phase. In the Ariadne \cite{Lonnblad:1992tz} dipole parton shower, which is used for final state radiation in \dipsy, minimization of string length is handled as a correction to the shower evolution, \ie on the perturbative level. Instead of emitting a gluon, and thus split into two new dipoles, a dipole $(1,2)$ can connect with another, colour compatible, dipole $(3,4)$, in the so-called swing mechanism, with the relative probability to reconnect:

\begin{equation}
	\label{eq:swing}
	\frac{dP}{d\rho} = \frac{(\vec{p}_1+\vec{p}_2)^2(\vec{p}_3+\vec{p}_4)^2}
  {(\vec{p}_1+\vec{p}_4)^2(\vec{p}_3+\vec{p}_2)^2}.
\end{equation}

Both $\lambda$-minimization and the swing mechanism are able to describe $\left<p_\perp\right>(N_{ch})$.

Junction topologies are notoriously difficult to handle kinematically. A recent attempt was implemented in \pytppp \cite{Christiansen:2015yqa}, in which larger junctions are broken down into smaller junctions, which are easier to handle. Junctions are usually thought to carry a natural baryon number,  but there are indications that this predicts too many baryons as compared to data. 
Higher multiplets can conveniently be described by the secondary Casimir operator ($C_2$) and the multiplicity of the multiplet. The effective string tension of the multiplet is then given by $\frac{\tilde{\kappa}}{\kappa} = \frac{C_2(\text{multiplet})}{C_2(\text{triplet})}$.

Since parameters for mass surpression in the string hadronisation model are related by equation (\ref{eq:dpdpt}), one can calculate how the parameters surpressing strange and baryon production\footnote{In ref. \cite{Bierlich:2015rha} we consider also spin suppresion, which is omitted here for brevity.}. When string tension increases, one produces more strange quarks in string breakings, as well as more diquarks, leading to baryons.

\section{Models for string overlap}
\label{sec:models}
To estimate the overlap region of strings, one needs a model for the transverse structure of the proton. Going slightly beyond $\lambda$-measures, we introduce here a toy model approach, which can be superimposed on \eg a \pytppp event, where strings are placed randomly in impact parameter space of the two protons. When a string breaks, it is checked if there are any other strings overlapping the breaking point in rapidity. If so, the two strings are decided to overlap if:
\begin{equation}
	4 \alpha^2 > (\sqrt{r_1}\cos \theta_1 - \sqrt{r_2} \cos \theta_2)^2 + (\sqrt{r_1}\sin \theta_1 - \sqrt{r_2} \sin \theta_2)^2, 
\end{equation}
where $\alpha$ is a parameter giving the ratio of string to proton transverse radius, and $r_i,\theta_i$ are random points on the unit circle. This approach gives qualitatively what is expected -- strings will overlap more in dense environments -- but the modelling of the proton transverse structure is clearly \textit{ad hoc}.

A more precise picture is given by the \dipsy model\cite{Flensburg:2011kk}. Here Fock states of the protons are built up from an evolution in impact parameter space and rapidity, giving a dynamic picture of the proton transverse structure.
\FIGURE[t]{
\centering
\includegraphics[width=0.45\textwidth]{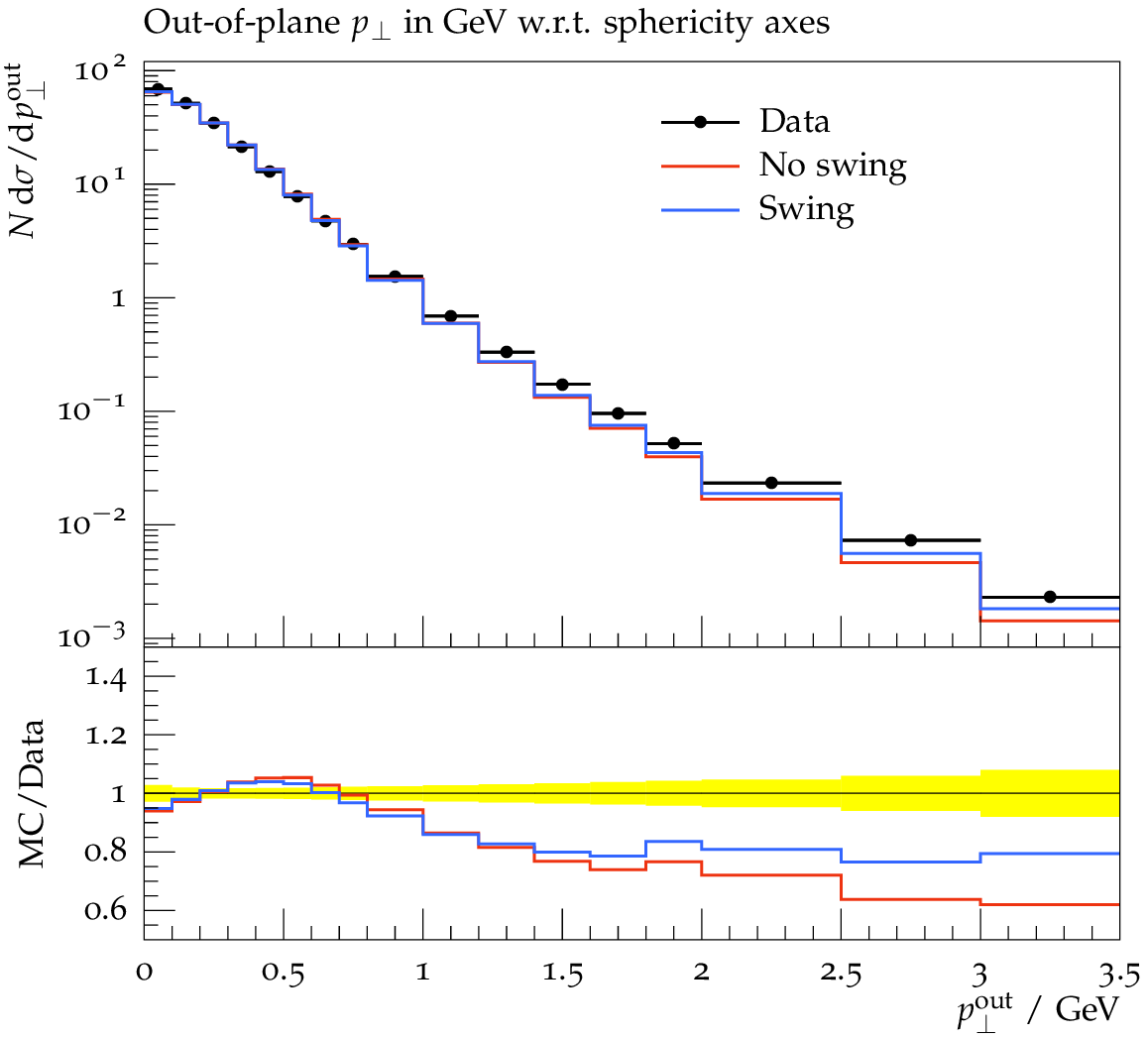}
\includegraphics[width=0.5\linewidth,angle=0]{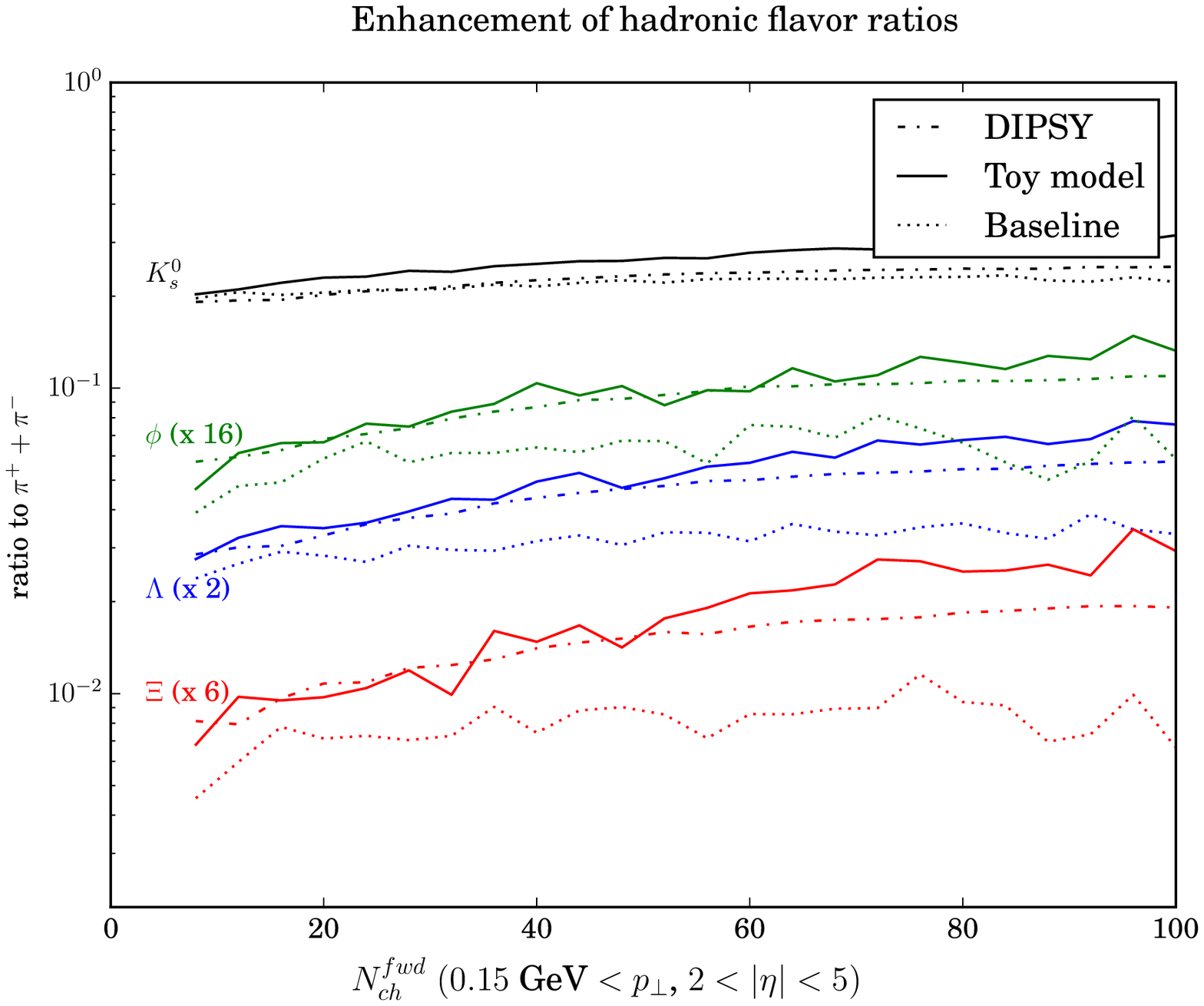}
\caption{(Left) The event shape observables $p_\perp^{out}$ recorded by DELPHI, compared to \dipsy/Ariadne with and without swing. A slight improvement is seen with swing. (Right) Integrated yield ratios for strange and multi-strange hadrons to pions. The dash-dotted line is \dipsy, the solid line the toy model, and the dotted is normal string hadronisation as a baseline.}
\label{fig:delphi}
}

In both the toy model and the full \dipsy model, overlap of strings has to be mapped to a multiplet structure as shown in figure \ref{fig:randomwalk}. In the full \dipsy model, this is done by perfoming a random walk, with all states weighted with the multiplicity of the multiplet. In the toy model an average is used.

\section{Final states}
\label{sec:data}
Even the simplest models of color reconnection should have some effects in $e^+e^-$. In figure \ref{fig:delphi}, we show the event shape observable $p_\perp^{out}$, as recorded by DELPHI \cite{Abreu:1996na}. The swing requires two additional emissions beyond the initial $Z\rightarrow q\bar{q}$ splitting. Since $p_{\perp,out}$ also requires two emissions, a small effect is thus expected and observed. We see that description is slightly improved by the swing, and more importantly, the difference already at this level, allows for validation using LEP data.

Since rope and color reconnection effects will rise with increasing event activity in $pp$, it is desirable to use a discriminator for event activity which can both be measured in experiments and be sensibly reproduced theoretically. Here we use the number of charged particles in the forward ($2 < |\eta| < 5$) direction. Measurements have been made of several of these observables already, and in \cite{Adam:2016emw} comparison to full DIPSY was shown, giving good agreement with all hadron species containing strange quarks\footnote{Since the measurements are not yet published in a form where direct comparison can be made, we only carry out a qualitative comparison.}.

In figure \ref{fig:delphi} we see a prediction for the $K^0_s/\pi$, $\Lambda/\pi$, $\phi/\pi$ and $\Xi/\pi$ ratios as a function of forward multiplicity, calculated with the toy model and full \dipsy for comparison. We see that both models show an increase of all species as a function of multiplicity, with the toy model having a slightly stronger rise. The $\phi/\pi$-ratio has not yet been measured, but all others were shown to compare well to data in ref. \cite{Adam:2016emw}. It is, however, paticularly interesting to measure the $\phi$. In models based on string hadronisation, the production of $\phi$ hadrons will be doubly enhanced compared to \eg $K^0_s$, while in models based on mass difference of hadrons, it will not. The increase of the double ratio $\phi/K^0_s$, is thus a strong prediction of the model.

It was shown by Ortiz \etal \cite{Ortiz:2013yxa} that even simple color reconnection can produce flow-like patterns in baryon/meson ratios as function of $p_\perp$, as measured in \eg ref. \cite{Adam:2015qaa}. In figure \ref{fig:flowlike} we show how the junction mechanism implemented in \pythia produces the same characteristic rise of the ratio with increasing event multiplicity as seen in data, but much stronger than for colour reconnection.

\FIGURE[t]{
\centering
\includegraphics[width=0.45\textwidth,angle=0]{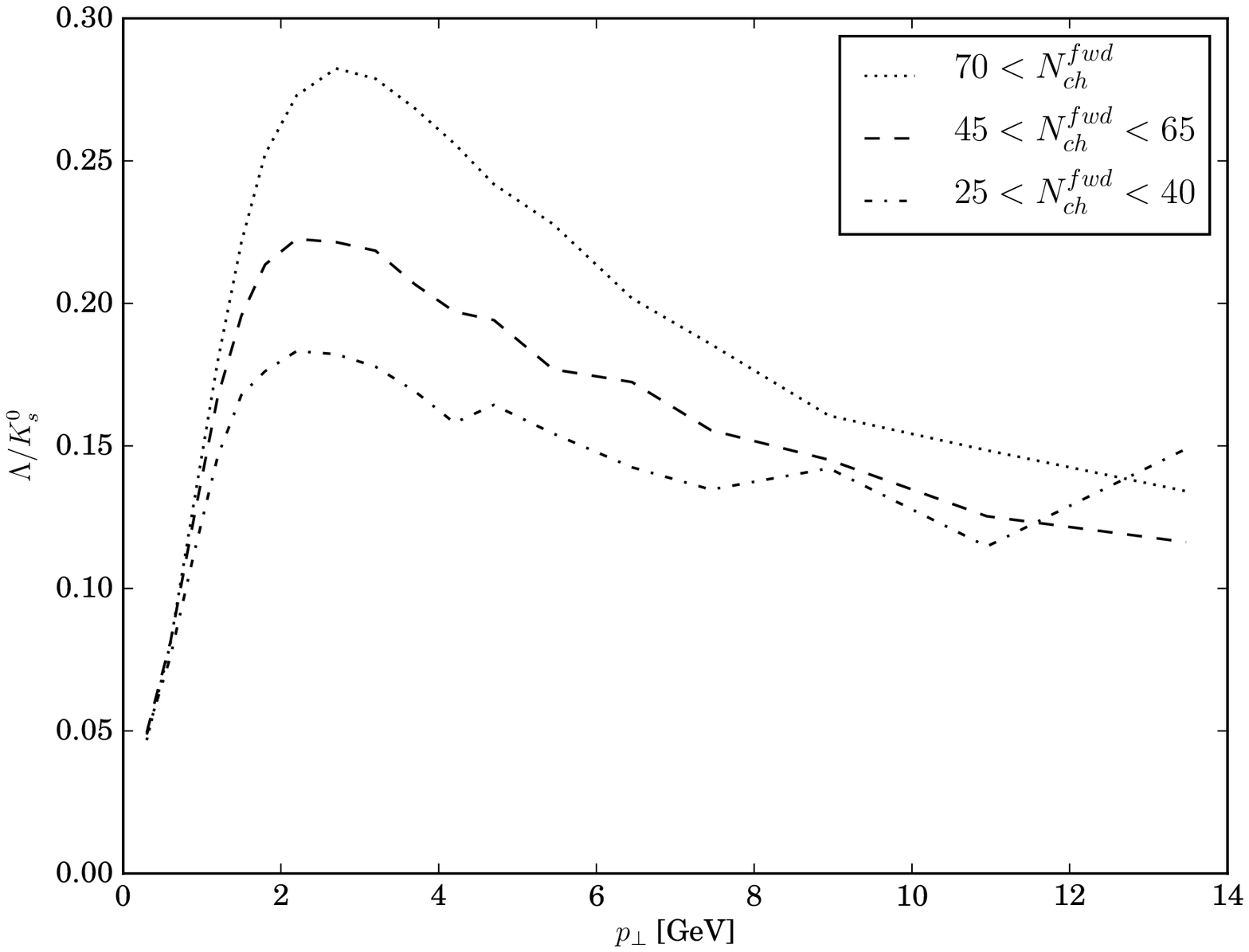}
\includegraphics[width=0.45\textwidth,angle=0]{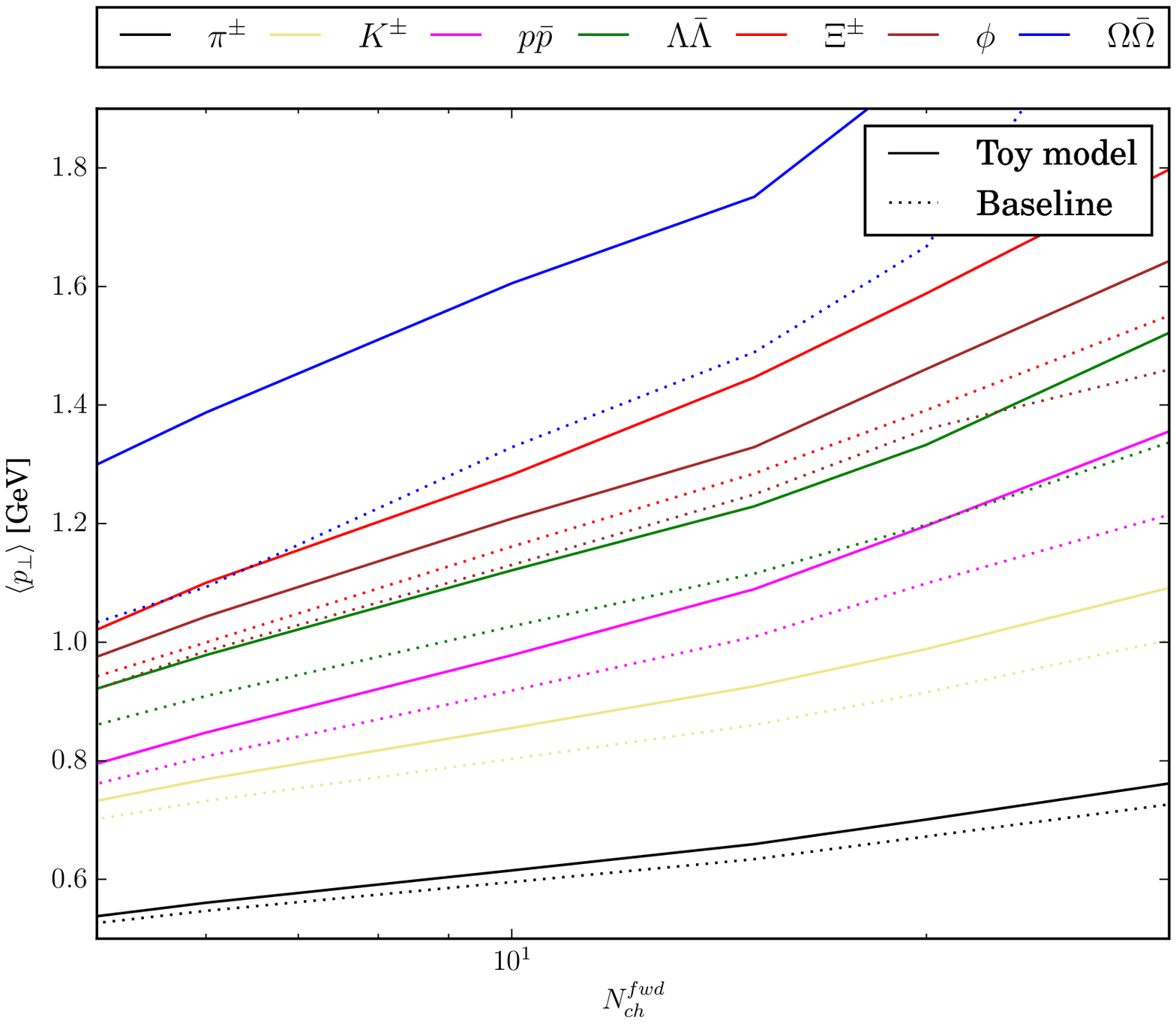}
\caption{(Left) Ratio of $\Lambda$ to $K^0_s$ as function of $p_\perp$, in bins of event multiplicity, calculated with \pythia, using the new junction implementation. Implementations of ropes which includes only simple or no junction handling are not shown. (Right) The $\left<p_\perp\right>(N_{ch})$ for identified particles, as calculated with the toy model.}
\label{fig:flowlike}
}

This is currently not reproduced by models such as \dipsy, which has only a very simplistic junction handling, or the toy model, which does not even include junctions.
Other observables sensitive to collective effects, and suitable for $pp$ exists. In ref. \cite{Ortiz:2015cma} it was suggested that $\left<p_\perp\right>$ for identified particles should be sensitive. New measurements done at ALICE \cite{Bianchi:2016szl} show a mass splitting of this observable, especially when is shown as function of event activity. As shown in figure \ref{fig:flowlike} this behavoir is reproduced qualitatively by ordinary colour reconnection, and enhanced by the toy model introduced here. For falsification of models of collectivity, including color reconnection and rope hadronisation, it is neccesary to allow for quantitative comparison of several models to data.

\bibliographystyle{JHEP}
\bibliography{refs}
\end{document}